\documentclass[conference]{IEEEtran}
\IEEEoverridecommandlockouts
\usepackage{cite}
\usepackage{amsmath,amssymb,amsfonts}
\usepackage{algorithmic}
\usepackage{graphicx}
\usepackage{siunitx}
\usepackage{textcomp}
\usepackage{xcolor}
\usepackage{hyperref}
\def\BibTeX{{\rm B\kern-.05em{\sc i\kern-.025em b}\kern-.08em
    T\kern-.1667em\lower.7ex\hbox{E}\kern-.125emX}}

    \def\SB[#1]{\textcolor{blue}{#1}}
\begin{document}

\title{Interpolation Filter Design for Sample Rate Independent Audio Effect RNNs
}

\author{
\IEEEauthorblockN{Alistair Carson\thanks{A. Carson is funded by the Scottish Graduate School of Arts and Humanities (SGSAH)}, Alec Wright and Stefan Bilbao}
\IEEEauthorblockA{\textit{Acoustics and Audio Group} \\
\textit{University of Edinburgh},
Edinburgh, UK \\
alistair.carson@ed.ac.uk, alec.wright@ed.ac.uk, sbilbao@ed.ac.uk}
}

% \and
% \and
% \and
% \and
% \and

\maketitle

\begin{abstract}
Recurrent neural networks (RNNs) are effective at emulating the non-linear, stateful behavior of analog guitar amplifiers and distortion effects. Unlike the case of direct circuit simulation, RNNs have a fixed sample rate encoded in their model weights, making the sample rate non-adjustable during inference. Recent work has proposed increasing the sample rate of RNNs at inference (oversampling) by increasing the feedback delay length in samples, using a fractional delay filter for non-integer conversions. Here, we investigate the task of lowering the sample rate at inference (undersampling), and propose using an extrapolation filter to approximate the required fractional signal advance. We consider two filter design methods and analyse the impact of filter order on audio quality. Our results show that the correct choice of filter can give high quality results for both oversampling and undersampling; however, in some cases the sample rate adjustment leads to unwanted artefacts in the output signal. We analyse these failure cases through linearised stability analysis, showing that they result from instability around a fixed point. This approach enables an informed prediction of suitable interpolation filters for a given RNN model before runtime. 
\end{abstract}

\begin{IEEEkeywords}
sample rate, recurrent neural network, audio effects
\end{IEEEkeywords}

\section{Introduction}
\label{sec:intro}

Virtual analog (VA) modeling refers to the digital emulation of analog audio effects and guitar amplifiers \cite{DAFXVAchapter}. The aim is to replace bulky, costly hardware with software---usually implemented as a plug-in for a digital audio workstation. Three main paradigms exist: \textit{white-box} approaches using circuit simulation methods \cite{Paiva12, Karjalainen06, Yeh2010}; \textit{black-box}, data-driven approaches \cite{Eichas:2016, Damskaag2018, Wright2020}; and hybrid \textit{grey-box} approaches such as differentiable DSP \cite{Nercessian:ICASSP2021, Wright2022, Carson2023, yeh2024ddspguitarampinterpretable}. For guitar amplifier and distortion emulation, a perceptually convincing black-box approach uses recurrent neural networks (RNNs) trained on paired input-output recordings of the specific device \cite{Wright2020}, sometimes conditioned on user-controls \cite{juvela2023end}. One limitation of RNNs compared to a white or grey box method is that the sample rate of the training data is implicitly encoded into the model weights and therefore not easily adjustable at inference. 

Ideally, any audio processing software should be able to operate at arbitrary sample rates---or at least the industry standard rates of  \SI{44.1}{\kHz},  \SI{48}{\kHz} and multiples thereof. One possibility is to resample the input signal to the desired rate, and then back to its original rate after processing. For real-time applications, however, this may add excessive CPU expense and/or latency. The performance will depend greatly on the choice of resampling filter, and therefore a detailed investigation into this is left to future work.

Here, we build on previous work \cite{Chowdhury2022, Carson2024} investigating modifications to the architecture of RNNs with the aim of a sample rate independent system. Other related work proposed sample rate independent convolutional neural networks (CNNs) for audio source separation \cite{Saito2022}, which may indeed be applicable to audio effect CNN models e.g. \cite{Damskaag2018, steinmetz2022, Comunita2022}, but here we focus solely on RNNs. Our contributions are:
\begin{itemize}
    \item to expand the non-integer oversampling experiments of \cite{Carson2024} with a larger set of candidate filters and  larger database of pre-trained LSTM models; 
    \item to explore the task of ``undersampling'' by a non-integer, proposing a means of achieving this by implementing a fractional signal advance in the state feedback loop; 
    \item to show that in many cases, the proposed interpolation filters can yield high quality results, but that in a few cases they severely degrade model output quality;
    \item to show that the cases where the interpolation/extrapolation filters fail can be analysed through a linearisation of the modified RNN around a fixed point. 
\end{itemize}
This paper is structured as follows: Sec. \ref{sec:problem_statement} introduces the problem; Sec. \ref{sec:srirnns} outlines sample rate independent RNNs and the proposed filter designs; Sec. \ref{sec:exp_details} contains the experimental details; Sec. \ref{sec:results} shows the results and Sec. \ref{sec:lin_analysis} investigates the results further through linear analysis. Sec. \ref{sec:conclusions} provides concluding remarks. Audio examples are available.\footnote{\scriptsize\url{https://a-carson.github.io/icassp25_srirnns/}}
\section{Problem Statement}
\label{sec:problem_statement}
Consider a continuous-time input audio signal $x(t)$ that has been sampled at a rate of $F_s$ to give $x_n$ where $n$ is an integer sample index. Here we consider RNN models of the form:
\begin{subequations}\label{eq:rnn_and_fc}
\begin{align}
    \mathbf{h}_n &= f\left(\mathbf{h}_{n-1}, x_n \right) \label{eq:rnn} \\ 
    y_n   &= g \left( \mathbf{h}_n, x_n \right),\label{eq:fc}
\end{align}
\end{subequations}
where $\mathbf{h}_n \in \mathbb{R}^{H \times 1}$ is the hidden state of length $H$ and $y_n$ is the output signal. This class of model has been extensively used in recent years for modelling guitar amplifiers and effects pedals\cite{Wright2020, Wright2019RNN, juvela2023end, cassidy2023perceptual}. In this work we consider $f$ to be a LSTM cell, and $g$ an affine transformation (which is by definition sample rate independent).

Consider the model \eqref{eq:rnn_and_fc} as pre-trained on audio with sample rate $F_s$ but at inference we wish to stream audio sampled at a different rate $F'_s$, such that the model takes the input signal $x'_n$ and produces the output signal $y'_n$. For an objective measure of quality we generate a target signal $\hat{y}_n$ which is $y_n$ resampled from $F_s$ to $F'_s$ using a DFT-based sample rate conversion \cite{Valimaki2023}, and measure the signal-to-noise ratio as:
\begin{eqnarray}\label{eq:snr}
    {\rm SNR} = \frac{\sum_{n=0}^{N-1} \hat{y}_n^2}{\sum_{n=0}^{N-1} \left( y'_n - \hat{y}_n \right)^2}
\end{eqnarray}
where $N$ is the duration of the input signal in samples. The objective here is to design a sample rate independent RNN which results in a maximum SNR.

%\vfill

\section{Sample rate independent RNNs}
\label{sec:srirnns}
Adjusting the inference sample rate of an RNN can be achieved by maintaining the same delay duration (in seconds) seen by the RNN during training \cite{Chowdhury2022, Carson2024}. The modified RNN operating at $F'_s$ can be defined as:
\begin{align}
    {\bf h}'_n &= f\left( {\bf h}'_{n-1 - \Delta}, x'_n \right)  \label{eq:srirnn_ideal} 
\end{align}
where $\Delta = F'_s/F_s - 1$ is the delay-line length adjustment in samples \cite{Chowdhury2022, Carson2024}. For non-integer conversion ratios, the state at non-integer time step $n-1-\Delta$ can be approximated with a fractional delay FIR filter:
\begin{align}
    {\bf h}'_{n-1 - \Delta} \approx \sum_{k=0}^{K} l_k {\bf h}'_{n-1-k} \label{eq:conv} 
\end{align}
where $K$ is the filter order and $l_k$ are the filter coefficients. 
% Note that ideally the delay $\Delta$ should be in a one-sample range centered around $K/2$
% for $0 < \Delta < 1$ and $K < 3$ a symmetric interpolation is possible here, but otherwise we are restricted to asymmetric interpolation to maintain causality. 

\subsection{Lagrange interpolation}\label{subsec:lagrange}
Lagrange interpolation is well known to be suitable for approximating a fractional delay \cite{Laakso1996, Kootsookos1996, Cain1994, Valimaki2000}, with the coefficients derived analytically \cite{Hermanowicz1992}:
\begin{eqnarray}\label{eq:lagrange_coeffs}
    l^{\rm lagr.}_k = \prod_{j=0, j \neq k}^{K} \frac{\Delta - j}{k - j}, \quad k = 0, \dots, K.
\end{eqnarray}
This design has the benefit of giving the exact phase response and unity gain at DC, i.e. $\sum_{k=0}^{K}l_{k}^{{\rm lagr.}} = 1$. If the desired delay is within a one-sample range and centered around $K/2$, the filter is maximally flat \cite{Laakso1996}. In this problem, however, we are restricted by the causality of \eqref{eq:srirnn_ideal} so we also consider non-centered designs. For $K=\{1, 3\}$ this filter was used for the task of oversampling ($\Delta > 0)$ in \cite{Carson2024}, with $K=3$ (non-centered) giving the best results across all the filters studied. Here we propose using a Lagrange extrapolation for approximation of a signal advance when $\Delta < 0$.

\subsection{Minimax design}\label{subsec:optimal}
Additionally we consider a \textit{minimax} design which minimises the L-infinity norm of the error magnitude over a desired bandwidth \cite{Laakso1996, Putnam1997}. Previous work applied this to a fractional delay problem \cite{Putnam1997} and showed the filter coefficients can be obtained via solution of a second order cone optimization \cite{Lobo1998}. Here, we set the bandwidth of optimization from zero to 0.25$F_s$ to cover the typical frequency range of guitar and bass effects. An additional constraint was imposed to find solutions where $\sum_{0}^{K} l^{\rm minimax}_k = 1$ to ensure unity gain at DC. The coefficients were obtained using MATLAB's \texttt{secondordercone} function.

\begin{figure*}[h!!!!]
\begin{minipage}[b]{.48\linewidth}
  \centering
  \centerline{\includegraphics[width=9.0cm, trim={0, 0.3cm, 0, 0.2cm}, clip]{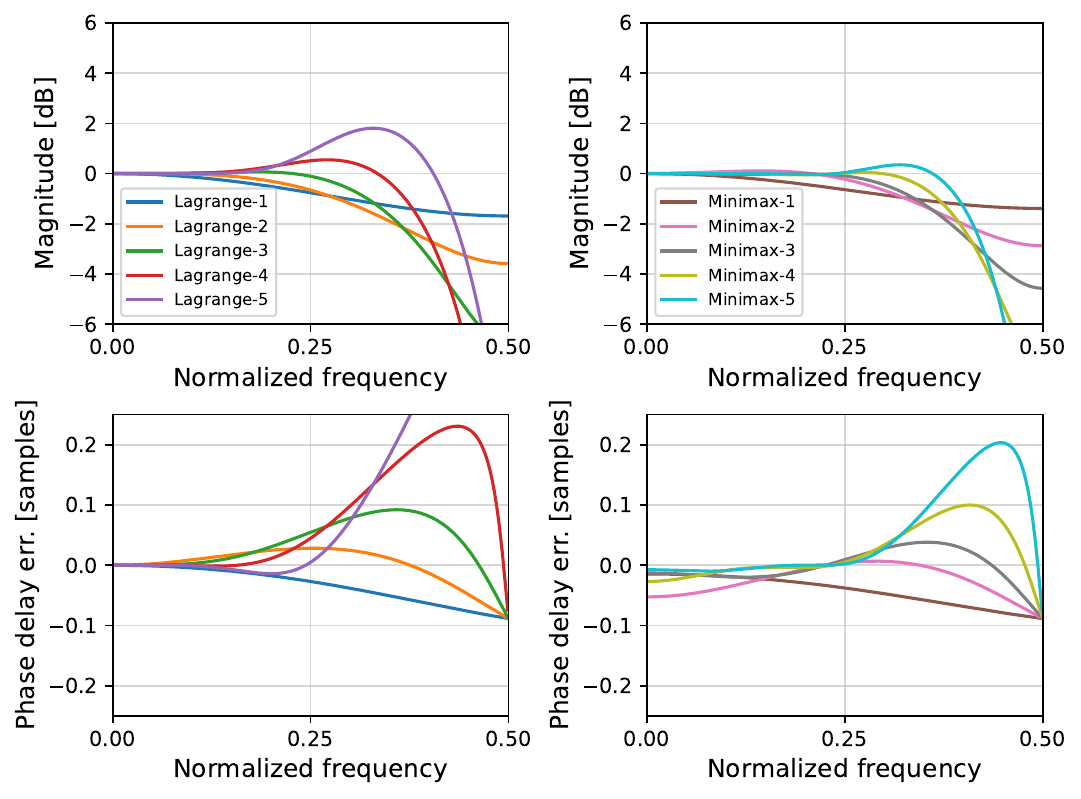}}
%  \vspace{1.5cm}
  \centerline{ \footnotesize{(a) Oversampling: $\Delta = 48/44.1-1$}}
\end{minipage}
\hfill
\begin{minipage}[b]{0.48\linewidth}
  \centering
  \centerline{\includegraphics[width=9.0cm, trim={0, 0.3cm, 0, 0.2cm}, clip]{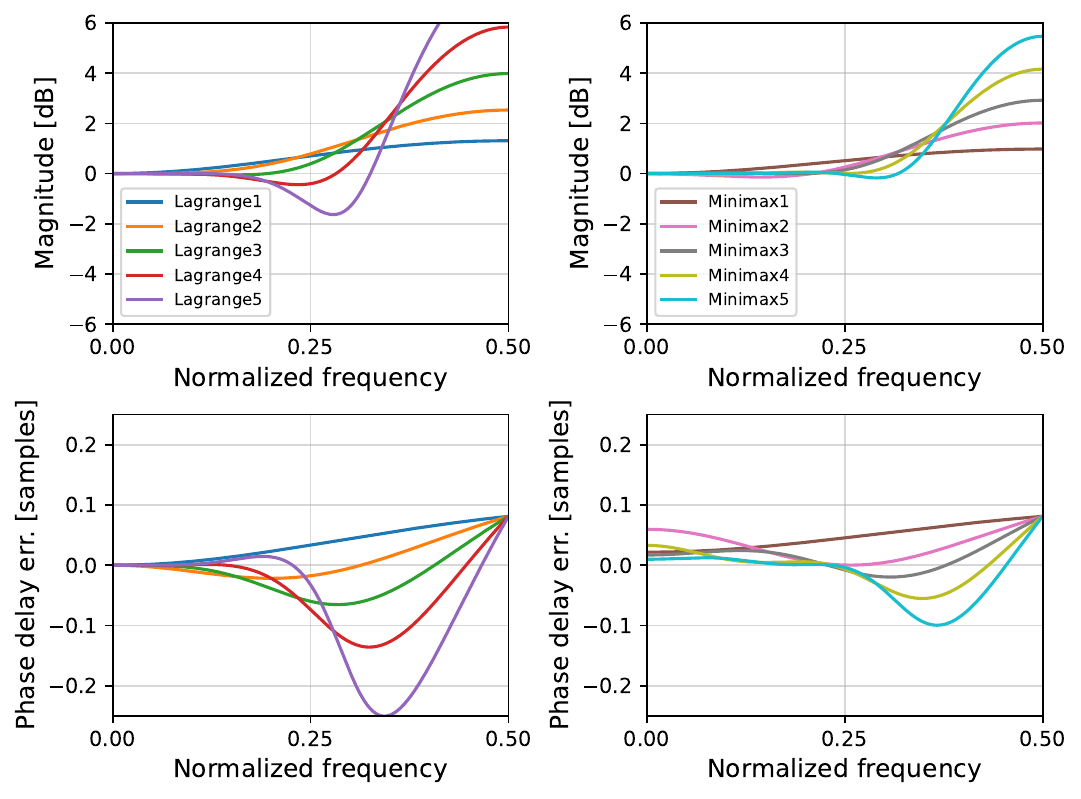}}
%  \vspace{1.5cm}
  \centerline{\footnotesize{(b) Undersampling: $\Delta = 44.1/48 -1$}}
\end{minipage}
\caption{Magnitude response (top) and phase delay error (bottom) of the candidate fractional delay filters for a) oversampling and b) undersampling.}
\label{fig:filters}
\end{figure*}

\section{Experiment details}\label{sec:exp_details}
\noindent {\bf Models:} all experiments were carried out on a set of 160 pre-trained LSTM models from the GuitarML Tonelibrary\footnote{\scriptsize\url{https://guitarml.com/tonelibrary/tonelib-pro.html}}. These models are user-created ``captures'' of various guitar amplifiers and effects pedals including distortion, overdrive and compressors. The models have the
same structure as Eq. \eqref{eq:rnn_and_fc} with $f$ being an LSTM cell with hidden size 40 giving $H=$ 80 states including the cell states.  The users are instructed to record their training data at $F_s=$ \SI{44.1}{\kHz}, so we assume the models have been trained correctly at this rate. Note that we have no information on model quality with respect to the original target analog system. \\

\noindent {\bf Sample rate conversion ratios:} we consider two common non-integer oversampling and undersampling ratios: $F'_s/F_s = \{160/147, \: 147/160\} $, giving operating sample rates of $F'_s = \{48, \: 40.5\}$\SI{}{kHz} for $F_s=$ \SI{44.1}{\kHz} \\

\noindent {\bf Candidate filters:} the candidate filters were designed using the Lagrange and Minimax methods for orders $K=\{1, \dots, 5\}$ and fractional delays of $\Delta = \{48/44.1 - 1, 44.1/48 - 1\}$. The magnitude response and phase delay error of the filters can be seen in Fig. \ref{fig:filters}. Henceforth, the K$^{th}$ order filter designs will be referred to as Lagrange-K or Minimax-K.\\

\noindent {\bf Baselines:} we compare the results against two baselines: the ``na{\"i}ve'' method of no interpolation or extrapolation; and the state-trajectory network method (STN) \cite{Parker2019, Carson2024}.\\ 

\noindent {\bf Test signal:} the input test signal  $x$ was sixty seconds of guitar and bass direct-input recordings. Before measuring the SNR the first 44100 samples were truncated to remove transients. \\

\section{Results}\label{sec:results}

\begin{figure*}[h!!!!]
\begin{minipage}[b]{.48\linewidth}
  \centering
  \centerline{\includegraphics[width=8.0cm, trim={0, 0.3cm, 0, 0.25cm}, clip]{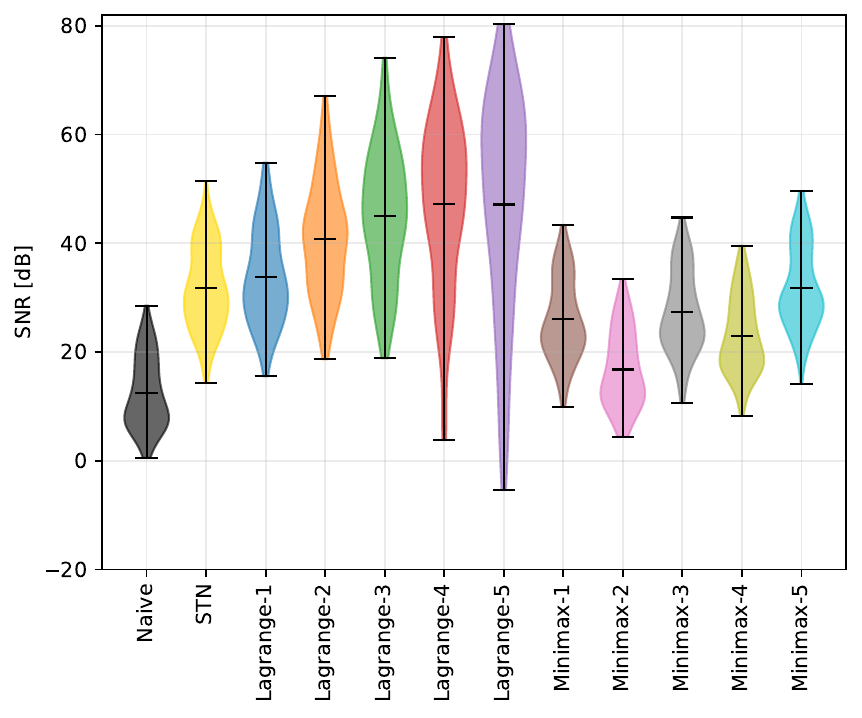}}
  \centerline{\footnotesize(a) Oversampling: \SI{44.1}{\kHz} to \SI{48}{\kHz}}
\end{minipage}
\hfill
\begin{minipage}[b]{0.48\linewidth}
  \centering
  \centerline{\includegraphics[width=8.0cm, trim={0, 0.3cm, 0, 0.25cm}, clip]{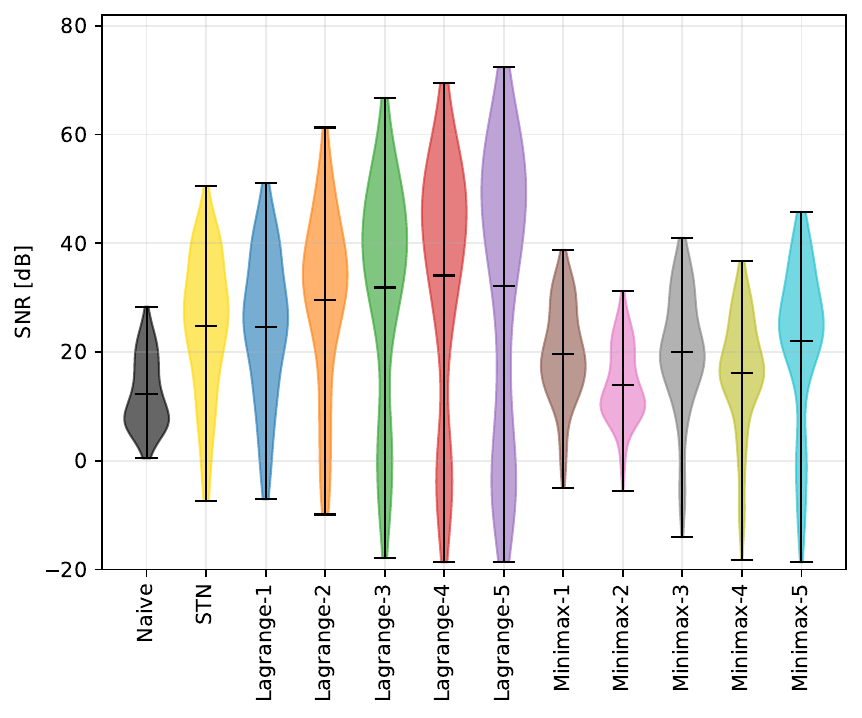}}
%  \vspace{1.5cm}
  \centerline{\footnotesize{(b) Undersampling:  \SI{44.1}{\kHz} to  \SI{40.5}{\kHz}}}
\end{minipage}
\caption{Violin plots showing the distribution of SNR results for interpolation/extrapolation with the different candidate filters. The horizontal lines indicate the minima, means and maxima respectively.}
\label{fig:violin}
\end{figure*}

% \begin{figure}[h!!!]\label{fig:pie_best}

% \begin{minipage}[b]{.48\linewidth}
%   \centering
%   \centerline{\includegraphics[width=4.5cm]{figs/exp1_pie_up_Best.pdf}}
% %  \vspace{1.5cm}
%   \centerline{\footnotesize{(a) 44.1kHz to 48kHz} }\medskip
% \end{minipage}
% \hfill
% \begin{minipage}[b]{0.48\linewidth}
%   \centering
%   \centerline{\includegraphics[width=4.5cm]{figs/exp1_pie_down_Best.pdf}}
% %  \vspace{1.5cm}
%   \centerline{\footnotesize{(b) 44.1kHz to 40.5kHz} }\medskip
% \end{minipage}
% %
% \caption{Number of cases where a given method ``wins'' (results in the best SNR across all filters and baselines) out of the 161 checkpoints.}
% %
% \end{figure}

% \begin{figure}[h!!!!!!!!!!!!!!]

% \begin{minipage}[b]{.48\linewidth}
%   \centering
%   \centerline{\includegraphics[width=4.5cm]{figs/exp1_pie_up_Worst.pdf}}
% %  \vspace{1.5cm}
%   \centerline{\footnotesize{(a) 44.1kHz to 48kHz} }\medskip
% \end{minipage}
% \hfill
% \begin{minipage}[b]{0.48\linewidth}\label{fig:pie_worst}
%   \centering
%   \centerline{\includegraphics[width=4.5cm]{figs/exp1_pie_down_Worst.pdf}}
% %  \vspace{1.5cm}
%   \centerline{\footnotesize{(b) 44.1kHz to 40.5kHz} }\medskip
% \end{minipage}
% %
% \caption{Number of cases where a given method ``loses'' (results in the worst SNR across all filters and baselines) out of the 161 checkpoints.}
% \label{fig:res}
% %
% \end{figure}

The results of the oversampling experiment are shown in Fig. \ref{fig:violin}a). Using Lagrange interpolation, the mean SNR increases with $K$, but so does the spread of the distribution across models. Lagrange-5 performed best in 58.1\% of cases (with a SNR of up to \SI{80}{\decibel}) however in 8.1\% of cases it performed worse than the na{\"i}ve method with a SNR as low as \SI{-5}{\decibel}. We will refer to these cases as failures, to be further examined in Sec. \ref{sec:lin_analysis}. The Minimax method produces on average a lower SNR than its Lagrange counterpart, especially for even $K$. This discrepancy is likely caused by phase-error at DC and low frequencies, where much of the test-signal energy is concentrated. As shown in Fig. \ref{fig:filters}, Lagrange interpolation is exact in phase and magnitude at DC for all $K$, and the even order Minimax filters have a higher phase error at DC compared to the neighboring odd orders. For a single ``best'' filter choice across all models, Lagrange-3 appears to be a good compromise with a SNR ranging from \SI{19}{\decibel} to \SI{73}{\decibel}.

Fig. \ref{fig:violin}b) shows the results for undersampling. In general the mean, minimum and maximum SNR for all filters is lower than in the oversampling case. In some cases the higher order Lagrange filters provided a good result of SNR $>$ \SI{60}{\decibel}, but the minimum SNR for all methods is less than \SI{0}{\decibel}, indicating an extremely noisy output signal in these cases. In 3.1\% of cases the na{\"i}ve method of no interpolation gave the best SNR and therefore none of the proposed methods are suitable for undersampling those models. We can conclude that there is no single filter for undersampling that will give reasonable results across all models. In the following section, we show that linearised analysis of the target system can help identify which filters are likely to fail, allowing the user to rule these out before experimenting at run-time.

\section{Linearised analysis}\label{sec:lin_analysis}
The results in Sec. \ref{sec:results} show that the best choice of interpolation filter has a strong dependence on the target system, and a poor choice can severely degrade output quality. Here we show that linearised analysis of the RNN \cite{sussillo2013opening} can help identify which filters are likely to cause a failure. 

Consider the modified RNN with delay-line interpolation \eqref{eq:srirnn_ideal} under zero input conditions $x_n = 0$. We can linearise around some fixed point ${\bf h} = {\bf a}$ through a Taylor expansion:
\begin{equation}\label{eq:taylor_srinn}
    {\bf h}_n \approx f({\bf a}) +   {\bf J_{a}} \cdot \left(\sum_{k=0}^{K} l_k {\bf h}_{n-1-k} - {\bf a}\right)
\end{equation}
where ${\bf J_{a}} = \nabla f \left( {\bf a} \right)^{T}$ is the Jacobian matrix. This can then be rewritten as a one-step state space system:
\begin{eqnarray}\label{eq:nice_state_space}
    {\bf v}_n = {\bf A} {\bf v}_{n-1} + {\bf b}
\end{eqnarray}
where ${\bf v} \in \mathbb{R}^{H\cdot (K+1) \times 1}$. Defining vector ${\bf l}=[l_0, \dots, l_K]$, we have block matrices:
\begin{equation}
{\bf A} =
\arraycolsep=1.0pt\def\arraystretch{1}
\left[
\begin{array}{c}
    \begin{array}{cc}
    {\bf l} \otimes {\bf J_{a}}
    \end{array} \\
    \hline
    \begin{array}{c|c}
    {\bf I}_{(HK) \times (HK)} & {\bf 0}_{(HK) \times H} \\
    \end{array}
\end{array}
\right]
, \:
{\bf b} = 
\arraycolsep=1.4pt\def\arraystretch{1}
\left[\begin{array}{c}
    \begin{array}{c}
    f({\bf a}) - {\bf J_{a}} {\bf a}
    \end{array} \\
    \hline
    \begin{array}{c}
    {\bf 0}_{(HK) \times H} \\
    \end{array}
\end{array}\right]
\end{equation}
where $\otimes$ is the Kronecker product and ${\bf I}$, $\bf{0}$ are identity and zeros matrices respectively. Note that for $K=0$ (no interpolation), ${\bf A} \equiv {\bf J_{a}}$. System \eqref{eq:nice_state_space} can then be analysed by examining the pole locations, given by:
\begin{equation}
    {\bf z}_{p} = {\rm eig}\left({\bf A}\right)
\end{equation}
In general we will have $H\cdot (K+1)$ non-unique eigenvalues, therefore the order of interpolation increases the number of poles in the system. For stability around this fixed point we require $\max \left( \left| {\bf z}_{p} \right| \right) \leq 1.0$.

\subsection{Experiment}
We approximated a fixed point for a given model by initialising the unmodified system with zero state and zero input, running for 10k samples and taking ${\bf a}$ as the time-averaged states over the last 1k samples. We then computed ${\bf z}_{p}$ for all model and filter combinations and predicted if each case would be either stable or unstable, depending on the pole locations. For each case the corresponding empirical result in Sec. \ref{sec:results} was labelled as either successful if the proposed methods gave a positive increase in SNR (relative to no interpolation) or a failure otherwise. This was repeated for both oversampling and undersampling to give 3200 experiments in total. The results can be seen in Table \ref{tab:stability}. The results show a high correlation between the stability prediction and the empirical results: in 97.8\% of cases ($(2868 + 261)/3200$) the linear analysis correctly predicts the binary empirical result.

\subsection{Case study}
\begin{figure}[htb]
\centering{\includegraphics[width=8.5cm, trim={0, 0.3cm, 0, 0.2cm}, clip]{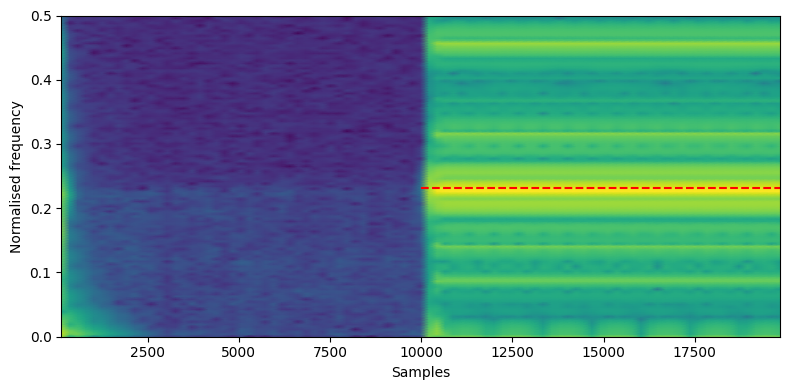}}
\caption{Output spectrogram of the MatchlessSC30 model checkpoint under zero-input conditions, initially with no interpolation/extrapolation in the delay line. At 10k samples, linear extrapolation is enabled, causing the model to self-oscillate. The red dashed line shows the unstable pole angle (see Fig. \ref{fig:pz}).}
\label{fig:spec}
\end{figure}

\begin{figure}[h]

\begin{minipage}[b]{.48\linewidth}
  \centering
  \centerline{\includegraphics[width=4.0cm, trim={0, 0.5cm, 0, 0.6cm}, clip]{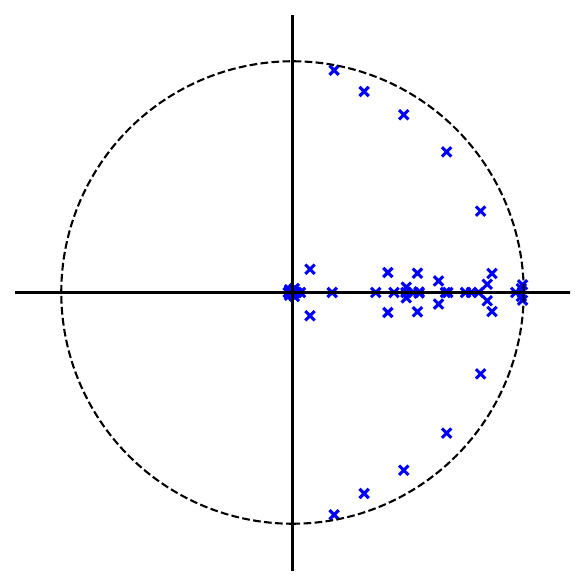}}
%  \vspace{1.5cm}
  \centerline{\footnotesize{(a)}}\medskip
\end{minipage}
\hfill
\begin{minipage}[b]{0.48\linewidth}
  \centering
  \centerline{\includegraphics[width=4.0cm, trim={0, 0.5cm, 0, 0.6cm}, clip]{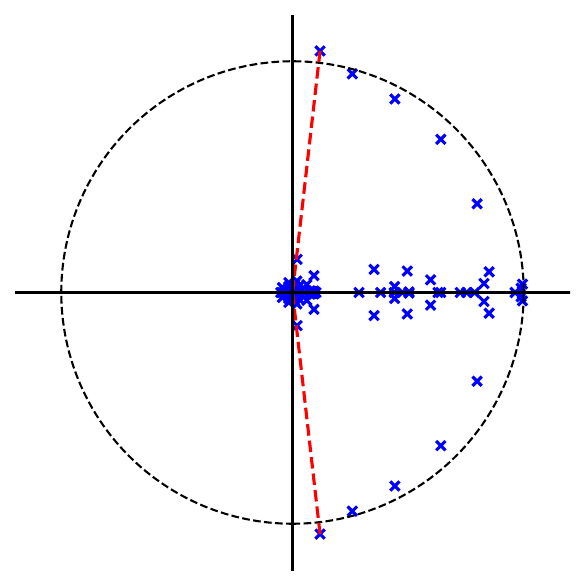}}
%  \vspace{1.5cm}
  \centerline{\footnotesize{(b)}}\medskip
\end{minipage}
\vspace{-0.4cm}
\caption{Poles of the linearised MatchlessSC30 model a) as trained (no interpolation) and b) with linear extrapolation in the delay-line. The black dashed line represents the unit circle. The red dashed lines show the pole angle of the unstable conjugate pair.}
\label{fig:pz}
\end{figure}

Fig. \ref{fig:spec} shows an example of a model that fails when linear extrapolation (Lagrange-1) is used to undersample the LSTM. The spectrogram shows high energy ringing at a frequency of $\sim0.22F_s$. Fig. \ref{fig:pz} shows the linearised analysis for the same system. The extrapolation has caused one conjugate pair of poles to leave the unit circle, and the pole-angle corresponds to the most prominent peak in the output spectrum. Referring back to Fig. \ref{fig:filters}(b) (top-left), this can be attributed to the high-shelf behaviour of linear extrapolation.

\begin{table}[h]
    \caption{Contingency table of stability analysis and empirical results}
    \centering
    \begin{tabular}{|c|cc|c|}
    \hline
      & Stable$\dagger$ & Unstable & Total \\ \hline
    Success$^*$ & 2868 & 55 & 2923 \\ 
    Failure & 16 & 261 & 277 \\ \hline
    Total & 2884 & 316 &  3200 \\ \hline
    \multicolumn{4}{l}{$\dagger$ $\: \max \left( \left| {\bf z}_{p} \right| \right) \leq 1.0$} \\
    \multicolumn{4}{l}{$^*$ positive increase in SNR compared to no interpolation.}
    \end{tabular}\\
    \label{tab:stability}
\end{table}

\section{Conclusions and further work}\label{sec:conclusions}
This paper explored interpolation-based methods for adjusting the sample rate of RNNs to allow audio processing at an inference sample rate that differs from training. This builds on prior work that proposed implementing a fractional delay filter in the state feedback loop for the task of non-integer oversampling. Here we extended this method to the inverse task of undersampling, and proposed achieving this by  approximating a fractional signal advance in the feedback loop. We considered Lagrange and minimax FIR filter designs, and evaluated the performance of the filters on 160 pre-trained LSTM models of various guitar amplifiers and distortion effects. The results showed that the best choice of filter was highly dependent on the effect-specific weights of the LSTM model. A good choice of filter may give up to \SI{80}{\decibel} SNR when oversampling or up to \SI{73}{\decibel} when undersampling. However, for certain models a poor choice of filter can result in poorer quality than if no interpolation was used. We showed that these failure cases can be predicted using linearised analysis of the original RNN around a fixed point. In future we will investigate model-specific optimal filter design, using the linearised analysis used to enforce stability as a design constraint. 
Alternatively, this analysis may enable sample rate independence by adjusting the weights of the network, thus avoiding the need for interpolation or extrapolation entirely.

% \section{TODO}
% \begin{itemize}
%     \item More references on background, resampling and linear analysis.
%     \item Does intro have enough context before presenting contributions?
%     \item Figure showing modified RNN structure? 
%     \item Clarify language about models/checkpoints/systems
% \end{itemize}
\bibliographystyle{IEEEtran}
\bibliography{refs}

\vspace{12pt}
\color{red}

\end{document}